\documentclass{moriond}
\usepackage{comment}

\def\be{\begin{equation}}
\def\ee{\end{equation}}
\def\bea{\begin{eqnarray}}
\def\eea{\end{eqnarray}}

\newcommand{\Photo}{\includegraphics[height=35mm]{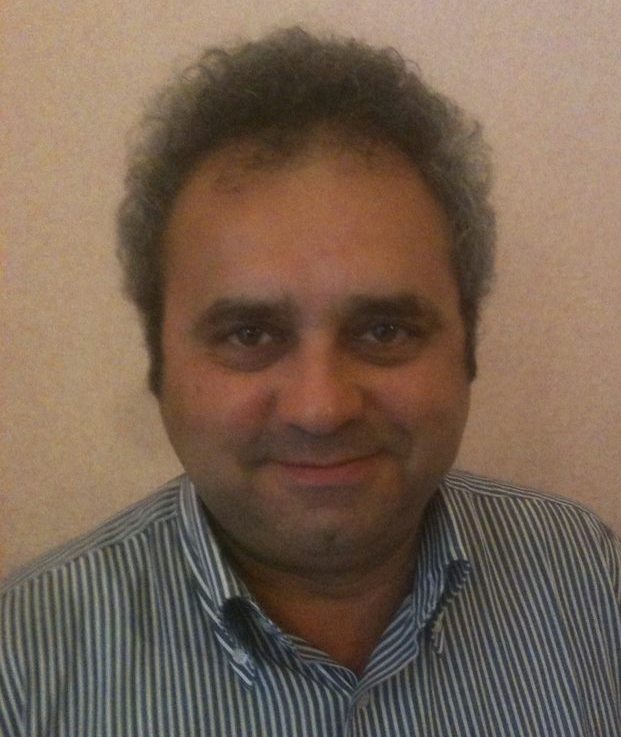}}

\begin{document}
\vspace*{4cm}
\title{STABILITY OF THE EW VACUUM, HIGGS BOSON, AND NEW 
PHYSICS\,\,\footnote{based on work done in collaboration 
with E. Messina\,\cite{bm}}}

\author{ VINCENZO BRANCHINA }

\address{Department of Physics and Astronomy, University of Catania, 
and INFN, Sezione di Catania \\
via S. Sofia, 63, I-95123, Catania, Italy}

\maketitle\abstracts{
The possibility that the Standard Model (SM) is valid up 
to the Planck scale $M_P$, i.e. that new physics occurs 
only around $M_P$, is nowadays largely explored. For a 
metastable EW vacuum, we show that new physics interactions 
can have a great impact on its lifetime, and, differently 
from previous analyses, they cannot be neglected. 
Therefore, contrary to usual believes, the stability 
phase diagram of the SM depends on new physics. This 
has far reaching consequences. Beyond SM theories can be 
tested against their prediction for the stability of the EW 
vacuum. Moreover, despite of some recent claims, 
higher precision measurements of the top and Higgs 
masses cannot provide any definite answer on the SM 
stability properties. Finally, doubts 
on Higgs inflation scenarios, all based on results
obtained neglecting new physics interactions, are also 
cast. }

\section{Stability phase diagram (new physics neglected)}

The Higgs effective potential $V_{eff}(\phi)$ bends down for values 
of $\phi$ larger than $v$, the location of the electroweak (EW) 
minimum (an instability due to top loop-corrections), and develops 
a new minimum at $\phi_{min}^{(2)} >> v$. Depending on 
Standard Model (SM) parameters, in particular on the top and 
Higgs masses, $M_t$ and $M_H$, the second minimum can 
be higher or lower than the EW one. In the first case the EW 
vacuum is stable, in the second one it is metastable and we have
to consider its lifetime $\tau$.

While several different scenarios for Beyond Standard 
Model (BSM) physics are considered, the possibility for the 
SM to be valid up to the Planck scale $M_P$,
meaning that new physics only occurs at scales 
around $M_P$, is not excluded and is the object of several
investigations. In the usual analysis, it is argued that 
new physics interactions at $M_P$ have no impact on the 
EW vacuum stability properties, and their presence is 
neglected\,\cite{isido,isiuno,isidue,giudice}. 

\begin{figure}\label{bounn}
\begin{minipage}{0.40\linewidth}
\centerline{\includegraphics[width=1.2\linewidth]{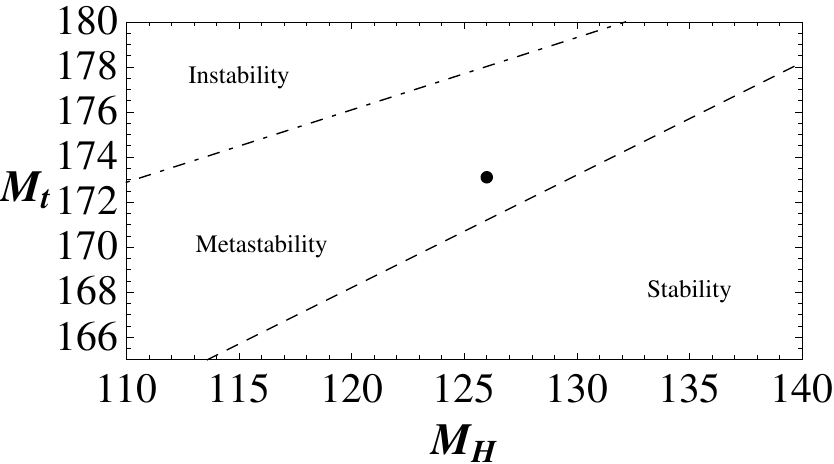}}
\end{minipage}
\hfill
\begin{minipage}{0.40\linewidth}
\centerline{\includegraphics[width=1.2\linewidth]{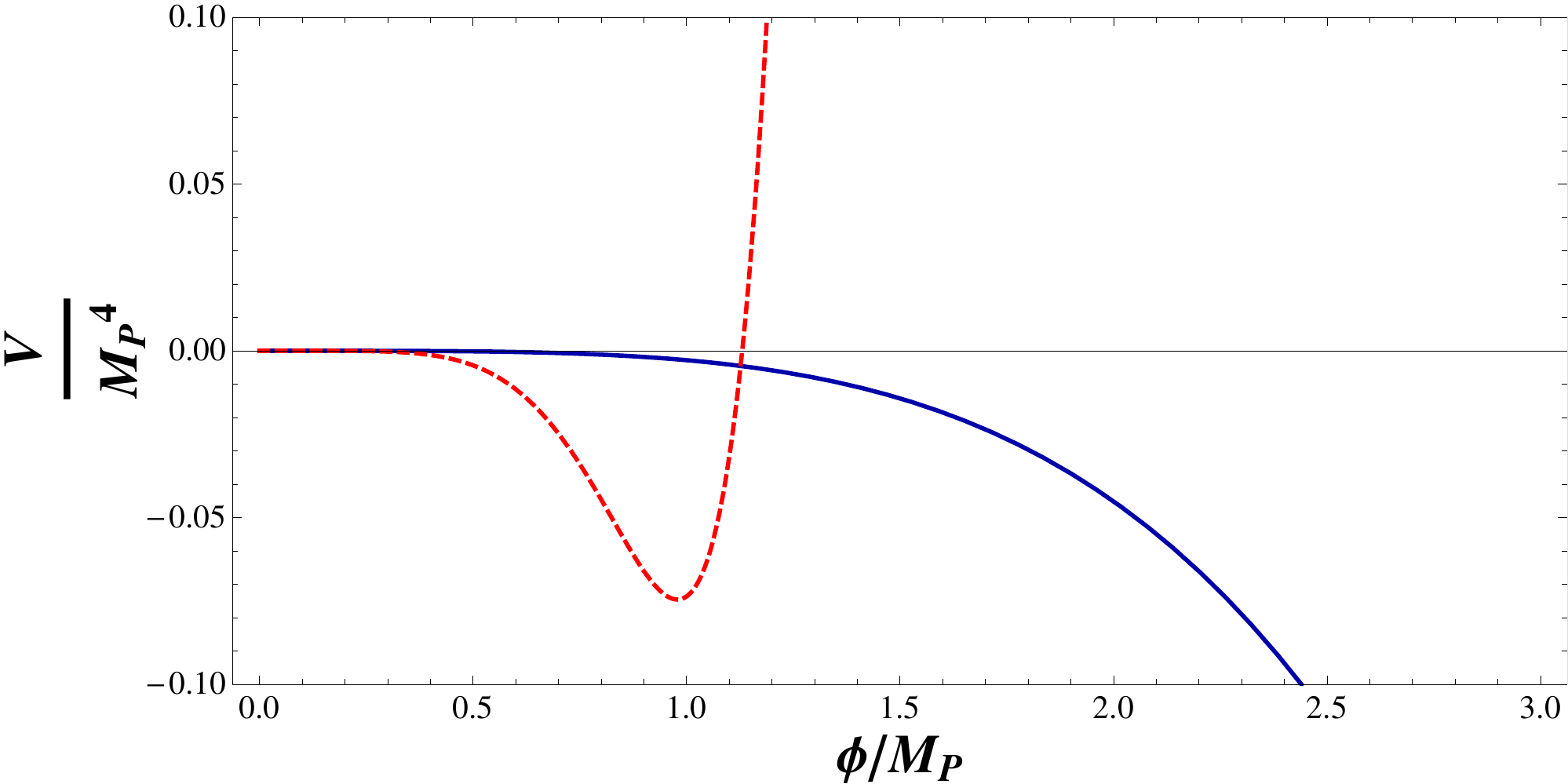}}
\end{minipage}
\caption[]{Left panel: Stability phase diagram obtained   
neglecting the presence of new interactions at the Planck scale. 
The $M_H-M_t$ plane is divided in three sectors,  
stability, metastability, and instability regions (see text). 
The dot indicates $M_H\sim 126$ GeV 
and $M_t\sim 173.1$ GeV.  Right panel: Effective 
potential $V_{eff}^{new}(\phi)$ (red line) in the presence of
the higher dimension operators $\phi^6$ and $\phi^8$, with  
$\lambda_6=-2$ and $\lambda_8=2.1$. For comparison, the blue 
line is for $V_{eff}(\phi)$ ($\lambda_6=0$ and $\lambda_8=0$).  }
\label{neweffpot}
\end{figure}

Under this assumption, the stability phase diagram of the SM in 
the $M_H-M_t$ plane turns out as shown in the left panel of 
fig.\ref{bounn}.
The plane is divided into three different sectors. 
An {\it absolute stability} region, where 
$V_{eff}(v) < V_{eff}(\phi_{min}^{(2)})$, a {\it  metastability} 
region, where $V_{eff}(\phi_{min}^{(2)}) < V_{eff}(v)$, but 
$\tau > T_U $, and an {\it instability} region, where 
$V_{eff}(\phi_{min}^{(2)}) < V_{eff}(v)$ and    
$\tau < T_U $, where $T_U$ is the age of the universe. 
The dashed line separates the stability and 
the metastability sectors and is obtained for 
$M_H$ and $M_t$ such that $V_{eff}(v) = 
V_{eff}(\phi_{min}^{(2)})$. The dashed-dotted line 
separates the metastability and the instability regions and is 
obtained for $M_H$ and $M_t$ such that $\tau = T_U $.

For $M_t \sim 173.1$ GeV and $M_H \sim 126$ GeV, the SM lies 
within the metastability region (black dot in the left panel 
of fig.\ref{bounn}). This observation leads to the 
so called  metastability scenario, that consists of the 
following proposal. Even though the EW vacuum is not 
the absolute minimum of $V_{eff}(\phi)$,  if $\tau > T_U$, 
our universe may well be sitting on such a metastable (false) 
vacuum.

There is another observation related to the position of the 
SM ``point'' in this figure. When the errors (not shown in
fig.\ref{bounn}) in the determination
of $M_H$ and $M_t$ are taken into account, it turns out that 
within 2.5-3 $\sigma$, 
the SM could be sitting on the dashed line, i.e. it could reach 
the stability region. This case is named ``critical'', as  
$\lambda$ at $M_P$ would reach the value 
$\lambda(M_P) \sim 0$, and also the beta function would be,
$\beta (\lambda(M_P)) \sim 0$. This ``near-criticality'' 
is considered by some 
authors the most important message from the experimental data 
on the Higgs boson\,\cite{giudice}. 

The above analysis, however, has some caveats. For the central 
values  $M_t \sim 173.1$ GeV and 
$M_H \sim 126$ GeV, for instance, the new minimum forms at 
$\phi_{min}^{(2)}$ much larger than $M_P$, 
$\phi_{min}^{(2)} \sim 10^{31}\, {\rm GeV}$ !
Despite of these (quite untrastable) results, it is argued that new 
physics at the Planck scale should stabilize the potential, 
bringing the new minimum around $M_P$, and that the computation 
of $\tau$ can be performed with the unmodified potential 
$V_{eff}(\phi)$, as the impact of new physics interactions 
should be negligible. 

Moreover, as the instability of the effective potential 
occurs for very large values of $\phi$, $V_{eff}(\phi)$ is 
well approximated by keeping 
only the quartic term\,\cite{sher2}, $V_{_{eff}}(\phi) 
\sim \frac{\lambda_{eff}(\phi)}{4} \phi^4$, where 
$\lambda_{eff}(\phi)$ depends on $\phi$ essentially as 
the running quartic coupling $\lambda(\mu)$ depends on the
running scale $\mu$. For large values of $\mu$, 
$\lambda(\mu)$ becomes negative and almost constant.
Therefore, $\tau$ is computed by considering first the 
bounce solution to the euclidean 
equation of motion for the potential $\frac{\lambda}{4}\phi^4$ 
with negative $\lambda$, and then taking into account the 
fluctuations around the bounce\,\cite{cole1,cole2}. 

In the following we show that new physics 
interactions at the Planck scale can dramatically change the lifetime 
of the metastable EW vacuum from $\tau >> T_U$ to $\tau << T_U$.

\section{Lifetime of the EW vacuum and new physics} 

The tunnelling rate $\Gamma$, inverse lifetime time $\tau$, 
is given by\,\cite{cole1,cole2} (for the sake of simplicity, 
we write the formula with the contribution of the 
scalar sector of the SM only, the inclusion of the 
other contributions being straightforward)
\begin{eqnarray}\label{tunneling}
\Gamma=\frac{1}{\tau} =	
T_U^{3}\,\frac{S[\phi_b]^2}{4\pi^2}
\left|\frac{ {\rm det'}\left[-\partial^2+V''(\phi_b)\right]}
{\mbox{det}\left[-\partial^2+V''(v)\right]}\right|^{-1/2} 
e^{-S[\phi_b]} 
\end{eqnarray}
where $\phi_b(r)$ is the $O(4)$ bounce solution to the euclidean 
equation of motion ($r=\sqrt {x_\mu x_\mu}$), 
$S[\phi_b]$ the action for the bounce, and 
$\left[-\partial^2+V''(\phi_b)\right]$ the fluctuation operator
around the bounce ($V''$ is the second derivative of $V$
with respect to $\phi$). The prime in the ${\rm det^{'}}$ 
means that the zero modes 
are excluded, and $\frac{S[\phi_b]^2}{4\pi^2}$ comes from the 
translational zero modes.

New physics interactions at the Planck scale appear as higher order 
operators multiplied by appropriate inverse powers of $M_P$. 
In order to study the possible impact of new physics 
at $M_P$, 
we now add to the SM Higgs potential two higher dimension operators 
$\phi^6$ and $\phi^8$, 
\begin{eqnarray}\label{newpot}
V(\phi)&=&\frac{\lambda}{4}\phi^4
+\frac{\lambda_6}{6}\frac{\phi^6}{M_P^2}
+\frac{\lambda_8}{8}\frac{\phi^8}{M_P^4}\,.
\end{eqnarray} 

In the right panel of fig.\ref{bounn}, the resulting effective 
potential $V_{eff}^{new}(\phi)$ (red line) for a specific 
example with natural values of $\lambda_6$ and $\lambda_8$,  
$\lambda_6=-2$ and $\lambda_8=2.1$, is plotted. 
For comparison, we also plot $V_{eff}(\phi)$ (blue line).
This example is well suited for our analysis. First of all, 
we have explicitly realized the stabilization 
of the effective potential around the Planck scale through the 
action of new physics operators as required in\,\cite{isido}
(see above). 
At the same time, we have a ``bona fide'' potential that we can 
use to check whether or not the usual 
assumption\,\cite{isido,isiuno,isidue,giudice} 
that in the evaluation of the EW vacuum lifetime new physics 
interactions can be neglected is correct.
As we shall see, they cannot be neglected and the stability 
phase diagram of fig.\ref{bounn} has to be revised.  

With $\lambda_6=0$ and $\lambda_8=0$, the euclidean equation 
of motion for the bounce can be solved analytically and we have 
\begin{eqnarray}\label{bounce1}
\phi_b^{(1)}(r) = 
\sqrt{\frac{2}{|\lambda|}}\frac{2R}{r^2+R^2}\,\,,
\end{eqnarray}
$R$ being the size of the bounce. The action is degenerate with 
$R$, \, $S[\phi_b^{(1)}]=\frac{\,8\,\pi^2}{3|\lambda|}$, the 
degeneracy being lifted by quantum fluctuations. The latters 
select only one bounce with given size, $R_1$. With 
$M_H \sim 126$ GeV and $M_t \sim 173.1$ GeV, 
$R_1\sim 8 \times 10^{-18}\,GeV^{-1}$.

With $\lambda_6 \neq 0$ and $\lambda_8 \neq0$, the euclidean 
equation of motion cannot be solved analytically. However, we 
can easily find the bounce numerically. Let us call 
$\phi_b^{(2)}(r)$ this solution. Due to the presence of the 
higher order terms, the degeneracy in this case is lifted 
already at the tree level, and the size is 
$R_2 \sim \frac{5.06}{M_P}$. For our scopes, it is important 
to note that, for $R >> 1/M_P$, the bounce 
$\phi_b^{(1)}(r)$ is also an approximate solution of the theory 
with non vanishing $\lambda_6$ and $\lambda_8$.  

If, according to the usual analysis, we now neglect 
the new physics interactions and compute the EW vacuum lifetime 
with $\lambda_6=0$ and $\lambda_8=0$, the tunnelling rate  
$\Gamma_0$ turns out to be 
\begin{eqnarray}\label{gamma1}
\Gamma_0=\frac{1}{\tau_{_0}}=\frac{1}{T_U} 
\left[\frac{S[\phi_b^{(1)}]^2}{4\pi^2}   \,
\frac{T_U^4}{R_{M}^4}\,e^{-S[\phi_b^{(1)}]}\right]\times 
\left[e^{-\Delta S_1}\right]\,,
\end{eqnarray}  
where $S[\phi_b^{(1)}]=\frac{\,8\,\pi^2}{3|\lambda|}$, $R_1$ 
and $T_U$ are as before, and $\Delta S_1$ is the loop 
contribution.

If, on the contrary, we take into account $\phi^6$ and $\phi^8$,
both $\phi_b^{(2)}(r)$ and the quasi solution $\phi_b^{(1)}(r)$ 
have to be considered, and for $\Gamma_{np}$ we get ($np$= new physics), 
\begin{eqnarray}\label{gamma2}
\Gamma_{np}=\frac{1}{\tau_{_{np}}}&=&\frac{1}{T_U} 
\left[\frac{S[\phi_b^{(1)}]^2}{4\pi^2}   \,
\frac{T_U^4}{{R_{1}}^4}\,e^{-S[\phi_b^{(1)}]}\right]\times 
\left[e^{-\Delta S_1}\right]\nonumber \\
&+& \frac{1}{T_U} 
\left[\frac{S[\phi_b^{(2)}]^2}{4\pi^2}   \,
\frac{T_U^4}{{R_2}^4}\,e^{-S[\phi_b^{(2)}]}\right]\times 
\left[e^{-\Delta S_2}\right]\,.
\end{eqnarray}
Inserting the obtained numerical values, 
$S[\phi_b^{(1)}]\sim 1833$	\,,\, 
$S[\phi_b^{(2)}]\sim 82$\,\,, $R_1\sim 8 \times 10^{-18}\,GeV^{-1}$,
and $R_2 ~\sim \frac{5}{M_P}$, even neglecting, for 
a moment, the one-loop $\Delta S_i$ contributions, we find 
from Eq.\,(\ref{gamma1}) (where $\lambda_6=0$, $\lambda_8=0$) and  
from Eq.\,(\ref{gamma2}) (where $\lambda_6=-2$, $\lambda_8=2.1$)
\begin{eqnarray}\label{tau1}
\tau_{_0}\sim 10^{555}\, T_U ~~~~~~~~~,~~~~~~~~~~
\tau_{_{np}}\sim 10^{-208}\, T_U\,\,.
\end{eqnarray}

Needless to say, Eq.\,(\ref{tau1}) clearly 
shows that new physics interactions at the Planck scale 
can have a dramatic impact on the EW vacuum lifetime. 
Moreover, from Eqs.\,(\ref{gamma2}) we see that the 
contribution to $\tau_{_{np}}$ coming from $\phi_b^{(1)}$ 
is exponentially suppressed. It is 
the bounce $\phi_b^{(2)}$, the one that we miss when we 
switch off the new physics interactions, that dominates! 

The reason for such an impact of new physics on $\tau_{_{np}}$ 
is easy to understand. New physics interactions appear in terms 
of higher dimensional operators, and we could naively expect 
their contribution to be suppressed. However, the tunnelling is 
a non-perturbative phenomenon. We first compute 
the bounce (tree level) and then the quantum fluctuations 
(loop corrections) on the top of it. The suppression in terms  
of inverse $M_P$ powers (power counting theorem) concerns 
the loop corrections, not the selection of the saddle point 
(tree level). The latter is intrinsically non-perturbative. 
In Eq.(\ref{newpot}) we have a new potential, and then a new 
saddle point. 

The inclusion of the $\Delta S_i$ does not change the above 
results significantly. For 
completeness, we write the values of $\tau_{_{0}}$ and  
$\tau_{_{np}}$ with the $\Delta S_i$ included : 
$\tau_{_{0}}\sim  10^{588}$ ~,~ $\tau_{_{np}}\sim  10^{-189}$. 

\section{Phenomenological consequences and conclusions}

The lifetime of the EW vacuum, as we have seen, strongly 
depends on new physics, and the stability phase diagram 
of fig.\ref{bounn} has to be revised. From the 
phenomenological point of view, this poses constraints on theories
beyond the SM. Any acceptable $UV$ completion of the $SM$
should not provide for $\tau$ results of the kind obtained
in the above example. In other words, our analysis provides a 
``BSM stability test'': a BSM theory is acceptable if 
it provides either a stable EW vacuum or a metastable one, 
with lifetime larger than the age of the universe. In 
the past it was thought that, given $M_H$ and $M_t$, the 
stability, metastability or instability of the EW vacuum 
could be established with no reference to the specific UV 
completion of the SM (stability 
phase diagram of fig.\ref{bounn}).  Clearly, our analysis 
can be repeated even when the new physics scale 
lies below the Planck scale (GUT scale, for instance).   

The ``near-criticality'' suggestion\,\cite{giudice}, 
$\lambda (M_P) \sim 0$ and  $\beta(\lambda (M_P)) \sim 0$,  
is also very much challenged by our results. The inclusion of 
new physics interactions can easily screw up these  
relations. The same is true for the Higgs inflation 
scenario of\,\,\cite{shapo1}, heavily based 
on the validity of the SM up to the Planck scale and on the 
criticality assumption\,\cite{shapo2}. 
Other Higgs inflation
scenarios, based on the possibility for the SM Higgs potential 
to develop a minimum at energies  $\sim 10^{16}$ GeV, 
where inflation could have started in a  
metastable state\,\cite{masina}, are also subject, for the same
reasons, to the same criticisms.

Finally, precision measurements of the 
top mass, that according to the phase diagram in 
fig.\ref{bounn} should tell us whether or not the SM moves
towards the stability line (the above discussed criticality), 
cannot give any answer to this question. As it should be clear 
by now, the knowledge of $M_t$ and $M_H$ is not sufficient 
to decide of the EW vacuum stability status.


\section*{References}

\begin{thebibliography}{99}
\bibitem{bm} V. Branchina, E. Messina, {\it Phys. Rev. Lett.} {\bf 111}, 
241801 (2013). 
\bibitem{isido} G. Isidori, G. Ridolfi, A. Strumia, {\it Nucl. Phys.} B 
{\bf 609}, 387 (2001).
\bibitem{isiuno} J. Elias-Miro, J. R. Espinosa, G. F. Giudice 
{\it et al.}, {\it Phys. Lett.} B {\bf 709}, 222 (2012). 
\bibitem{isidue} G. Degrassi, S. Di Vita, J. Elias-Miro {\it et al.}, 
{\it JHEP} {\bf 1208}, 098 (2012).
\bibitem{giudice} D. Buttazzo, G. Degrassi, P. P. Giardino et al., 
JHEP 1312 (2013) 089
\bibitem{sher2} M. Sher, {\it Phys. Lett.} B {\bf 317}, 159 (1993).
\bibitem{cole1} S. Coleman, {\it Phys. Rev.} D {\bf 15}, 2929 (1977). 
\bibitem{cole2} C. G. Callan, S. Coleman, {\it Phys. Rev.} D
{\bf 16}, 1762 (1977).
\bibitem{shapo1} F. L. Bezrukov and M. Shaposhnikov, {\it Phys. Lett.} 
B {\bf 659}, 703 (2008).
\bibitem{shapo2} F. Bezrukov, M. Shaposhnikov, arXiv:1403.6078. 
\bibitem{masina} I. Masina, A. Notari Phys.Rev. D85, 123506 (2012); 
I. Masina, arXiv:1403.5244. 
\end{thebibliography}
\end{document}